\documentstyle[aps,epsf,epsfig]{revtex}        

\begin{document}


\title{Nonequilibrium Superconductor-Normal Metal  Tunnel Contact and the Phonon Deficit Effect}

\author{Gurgen  Melkonyan$^1$\footnote{gmelkony@phy.ulaval.ca}, Helmut Kr\"oger$^1$\footnote{hkroger@phy.ulaval.ca}and}
\author{Armen M. Gulian$^2$}

\address{$^1$D\'epartement de Physique, Universit\'e Laval, Qu\'ebec, Qu\'ebec, Canada, G1K 7P4}
 
\address{$^2$USRA/US Naval Research Laboratory, Washington, DC 20375, USA }

\maketitle

\begin{abstract}
We consider tunnel microrefrigerators at low temperature. There
is a number of experimental studies performed on
microrefrigeration in tunneling superconductor--normal metal ($SN$) structures. Related to these experiments, only the electron
subsystem has been considered theoretically. Independently, the phonon deficit
effect  has been studied a while ago in superconductor-superconductor
tunnel junctions. It can be regarded as a possible prototype scheme for
superconducting microrefrigerators. We try to provide the missing link
between experiments on the $SN$ tunnel junction refrigerators and the theory
which includes microscopically phonons in combination with the mechanism of the phonon
deficit effect.

\end{abstract}



 A normal metal--superconductor tunnel contact can be used to cool the normal metal electrode and a membrane linked to it ~\cite{pekle}.  Many groups have investigated $SN$ tunnel contact as the prototype for
the tunnel contact microrefrigerator or electronic refrigeration ~\cite{caspek} device on a chip. All electron subsystem cooling experiments are based on Parmenter's idea ~\cite{parm} (or equivalently the Peltier effect for non--superconducting case)  to find a method of extracting the normal electrons, thermally excited from the gap edge region, into one of the superconducting electrodes in a superconductor--superconductor  ($SS^\prime$) tunnel contact.  The $SN$ contact  used  in these  experiments can be considered as the limiting case of  an $SS^\prime$ tunnel contact.

The phonon "deficit effect" was initially predicted as an effect accompanying the enhancement of superconductivity in a high--frequency electromagnetic field ~\cite{eliash}. It has widely  been studied~\cite{gulzhar} for a variety of different situations including refrigeration in the $SS^\prime$ contacts. 
Here we will consider the behavior of an $SN$ tunnel contact as a limiting case of an asymmetrical $SS^\prime$ tunnel contact.

As in~\cite{gulzhar} we consider a massive superconductor ($S_2$) with the BSC gap $\Delta_2$ and the critical temperature $T_{C_2}$ linked to a thin normal metal ($N$) or superconducting film with thickness $d \sim \xi_0 \sim v_F/ T$ by means of an oxide layer (here $v_F$ is the particle velocity at the Fermi level, $\xi_0$ is the coherence length). This system is linked to a thermostat at a temperature  $T$. The electron and hole  distribution functions in the massive superconductor are those at  equilibrium due to rapid diffusion reabsorption of the tunneling excitations. The distribution of electrons  ($n_\epsilon$) and holes ($n_{-\epsilon}$) in the thin film ($S_1$ or $N$) is described by the kinetic equation
\begin{equation}
\label{particle}
u_\epsilon\frac{\partial n_{\pm \epsilon}}{\partial t}=Q(n_{\pm \epsilon})+J(n_{\pm \epsilon},N_{\omega}).
\end{equation}   
 The explicit form of the tunnel source and electron--phonon collision integral operators are given in~\cite{gulzhar}, $u_\epsilon=|\epsilon|\theta(\epsilon^2-\Delta^2)/\sqrt{\epsilon^2-\Delta^2}$ is the BCS density of states, $\epsilon$ is the electron energy. Because the smallness of the   film ($S$ or $N$) thickness $d$   we assume that the phonons have a sufficient time to leave the film and they do not back influence  on the electrons in this thin film. At this condition we take $N_\omega= N_\omega^0$  ($N_\omega^0$ is the Bose distribution function for phonons) for $J(n_\epsilon,N_\omega)$ in equation ~(\ref{particle}) as in~\cite{gulzhar}.  If the system is in the stationary state then we can solve by iterations the equation~(\ref{particle})  and find stationary state solutions for particle and hole distribution functions.  The intensity of the phonon radiation at a frequency  $\omega_q$ in a range  $d\omega_q$ is equal to
\begin{equation}
\label{phonemmison}
\frac{d\dot{P}_{\omega_q}}{d\omega_q}=\frac{v  \lambda \omega_D \omega_q^3}{16\pi \epsilon_F u^3}I(N^0_{\omega_q}),
\end{equation} 
where ${ v}$ is the volume of the film, $\omega_D$, $\lambda$, $u$, $\epsilon_F$ are the Debye frequency,  the dimensionless electron--phonon interaction constant,  the sound velocity and the Fermi energy, respectively.  $I(N^0_{\omega_q})$ is the phonon--electron collision operator. There are  negative and  positive phonon fluxes ~\cite{gulzhar}  between the thin film and the thermal reservoir. The net phonon flux under certain conditions can be either negative or positive . 

\begin{figure}[btp]
\begin{center}\leavevmode
\includegraphics[width=0.95\linewidth]{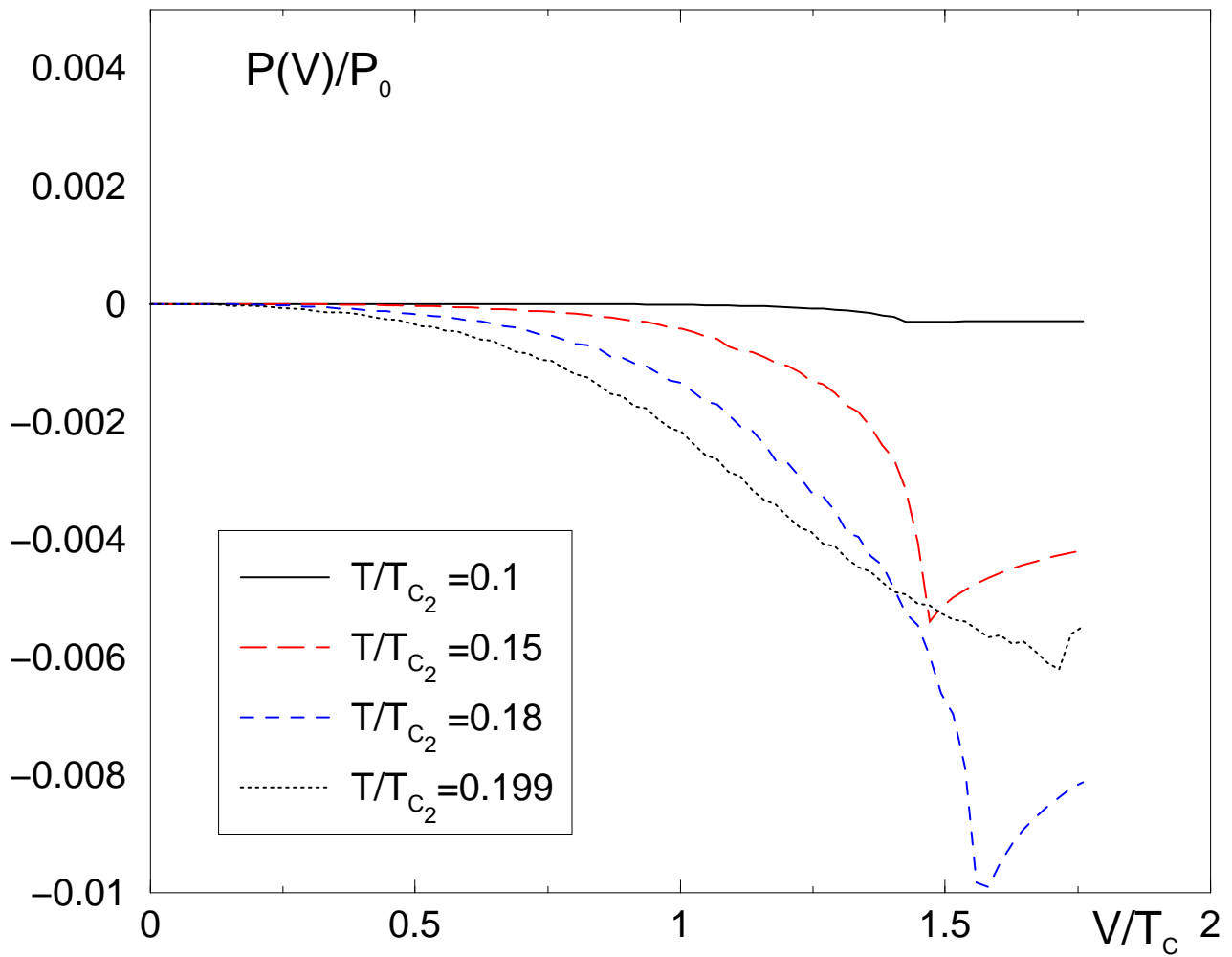}
\caption{Phonon absorption curves for different temperatures. $P_0={v }\lambda \omega_D T_{C_2}^5/(16\pi \epsilon_F u^3$). $T_{C_2}=1K$, $T_{C_1}=0.2K$, $\Delta_2(T)/T_{C_2}=1.759$ and  $\Delta_1(T)/T_{C_2}$ is equal to $0.34$, $0.29$,  $0.19$ and $0.04$ from top   to bottom of insert respectively.
}
\label{figone}\end{center}\end{figure}
Fig. ~\ref{figone} illustrates the results of our numerical simulation for the phonon emission (absorption) at different voltages ($V$)  for the $SS^\prime$ tunnel contact.  
 We can see that at low temperatures cooling capacity of the asymmetric $SS^\prime$ contact  is low and as $\Delta_1$ goes to zero its cooling capacity is increased.  Above the temperature $T>T_{C_1}$ the  cooling capacity of the tunnel contact decreases.  This means that the BCS peculiarities in the electrons density of states  plays a very important role  for the cooling process. In the case of the normal metal this picture  changes so  that there will be  almost only phonon absorption from the heat--bath at low temperatures and phonon emission to the heat--bath at higher temperatures.    From  the above  analysis it is clear how to optimize the parameters of the   $SS^\prime$ ($SN$) tunnel contact to get  suitable device for  low temperature microrefrigeration.
 
This work has been supported by NSERC Canada.  G. M. is grateful to the "Programme qu\'eb\'ecois de bourses d`excellence" of Qu\'ebec for the financial support. 

 

\begin{thebibliography}{9}
\bibitem{pekle} A.J. Manninen, M.M. Leivo and J.P. Pekola, Appl. Phys. Lett.  {\bf 70}  (1997)  1885.

\bibitem{caspek}  M.M. Leivo,  J.P. Pekola and D.V. Averin,  Appl. Phys. Lett., {\bf 68}  (1996) 1996; M.G. Castelano et al., Nuovo  Cimento  {\bf 19}  (1997) 1417.

\bibitem{parm}  R.H. Parmenter, Phys.  Rev. Lett. {\bf 7}  (1961)  274.  
\bibitem{eliash}  G.M. Eliashberg, JETP Lett.  {\bf 11}  (1970) 114. 
 \bibitem{gulzhar}  A.M. Gulian and G.F. Zharkov, Phys. Lett.  A   {\bf 80} (1980) 79;  JETP Lett.  {\bf 34} (1981) 153; A.M. Gulian, G.F. Zharkov and G.M. Sergoyan,  Kratkie Soobshcheniya po Fizike FIAN No. 10 (1985) 33  [{\it Sov. Phys. Lebedev Inst. Reports }  {No. 10} (1985) 38  ].
\end{thebibliography}
\end{document}